# Multiparameter Water Quality Monitoring System for Continuous Monitoring of Fresh Waters

Damir B. Krklješ, Goran V. Kitić, Csaba M. Petes, Slobodan S. Birgermajer, Jovana D. Stanojev, Branimir M. Bajac, Marko N. Panić, Vasa M. Radonić, Ilija D. Brčeski, Rok M. Štravs, Nikolina N. Janković, Jovan B. Matović

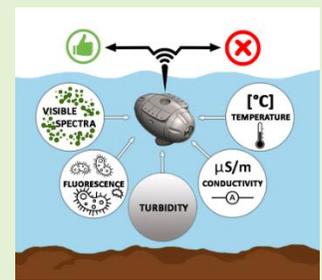

***Abstract*—** Due to the global water crisis there is a strong need for real-time water quality monitoring with high temporal and spatial resolution. This paper presents an economical multiparameter water quality monitoring system for continuous monitoring of fresh waters. It is based on a sensor node that integrates turbidity, temperature, and conductivity sensors, a miniature eighteen-channel spectrophotometer, and a sensor for the detection of thermotolerant coliforms, which is a major novelty of the system. Due to the influence of water impurities on the measurement of thermotolerant coliforms, a heuristic method has been developed to mitigate this effect. Moreover, the sensor is low power and with an integrated Long Range Wide Area Network module, it comprises a system that is wireless sensor network ready and can send data to a dedicated server. In addition, the system is submersible, capable of long-term field operation, and significantly cheaper in comparison to existing solutions. The purpose of the system is to give early warning of incidental pollution situations, thus enabling authorities to take action regarding further prevention of such occasions.

***Index Terms*—** Escherichia coli, sensor node, thermotolerant coliform bacteria, turbidity sensor, conductivity sensor, spectrophotometer, wireless sensor network.

## I. INTRODUCTION

WATER resources represent the basic source of life as well as a very important social and economic resource factor, whose quantitative and qualitative potentials are dramatically decreasing everywhere in the world [1]. The main causes of potential decline are urban wastewater, intensive agriculture, especially large farms, and industries. They either dump huge amounts of waste materials into watercourses and make the water unusable or require large investments in water purification, as well as broad cost-benefit analysis [2]. For instance, the steel industry typically requires 28 liters of water to produce 1 kilogram of steel [3], and the estimation of the paper industry consumption is between 300 and 2600 liters of water to produce a kilogram of paper [4]. Moreover, around 60% of artificial fertilizers end up in underground water that further emerges on the surface leading to excessive growth of algae and aquatic plants, which release toxic substances during

This work was supported by the Ministry of Education, Science and Technological Development of the Republic of Serbia and the Slovenian Government through the Eureka Project under Grant E!13044.

This work is supported in part through ANTARES project that has received funding from the European Union's Horizon 2020 research and innovation programme under Grant SGA-CSA. No. 739570 under FPA No. 664387. https://doi.org/10.3030/739570.

D. B. Krklješ is with the BioSense Institute, University of Novi Sad, Dr Zorana Đinđića 1, 21000 Novi Sad, Serbia (e-mail: dkrkljes@biosense.rs).

G. V. Kitić is with the BioSense Institute, University of Novi Sad, Dr Zorana Đinđića 1, 21000 Novi Sad, Serbia (e-mail: gkitic@biosense.rs).

C. M. Petes is with the BioSense Institute, University of Novi Sad, Dr Zorana Đinđića 1, 21000 Novi Sad, Serbia (e-mail: chaba@biosense.rs).

S. S. Birgermajer is with the BioSense Institute, University of Novi Sad, Dr Zorana Đinđića 1, 21000 Novi Sad, Serbia (e-mail: b.sloba@biosense.rs).

J. D. Stanojev is with the BioSense Institute, University of Novi Sad, Dr Zorana Đinđića 1, 21000 Novi Sad, Serbia (e-mail: jovana.stanojev@biosense.rs).

B. M. Bajac is with the BioSense Institute, University of Novi Sad, Dr Zorana Đinđića 1, 21000 Novi Sad, Serbia (e-mail: branimir.bajac@biosense.rs).

M. N. Panić is with the BioSense Institute, University of Novi Sad, Dr Zorana Đinđića 1, 21000 Novi Sad, Serbia (e-mail: panic@biosense.rs).

V. M. Radonić is with the BioSense Institute, University of Novi Sad, Dr Zorana Đinđića 1, 21000 Novi Sad, Serbia (e-mail: vasarad@biosense.rs).

I. D. Brčeski is with the Faculty of Chemistry, University of Belgrade, Studentski trg 12-16, 11158 Belgrade, Serbia (e-mail: ibrceski@chem.bg.ac.rs).

R. M. Štravs is with BIA d.o.o, Cesta v gorice 34, SI-1000 Ljubljana, Slovenia (e-mail: rok.stravs@bia.si).

N. N. Janković is with the BioSense Institute, University of Novi Sad, Dr Zorana Đinđića 1, 21000 Novi Sad, Serbia (e-mail: nikolina@biosense.rs).

J. B. Matović is with the BioSense Institute, University of Novi Sad, Dr Zorana Đinđića 1, 21000 Novi Sad, Serbia (e-mail: matovic.matovic@gmail.com).

decomposition [5]. Although not an [6]–[11] list, these several examples clearly indicate how endangered water sources are in terms of extensive consumption and pollution.

Especially sensitive waters are those classified of high and good status according to the European Water Framework Directive [12] due to their relative scarcity and difficulties to protect them. Besides anthropogenic influences and excessive intake of water during heavy rains and snow melting, one of the sources of their contamination is livestock feces. A strong indicator of fecal contamination is the presence of the thermotolerant coliform bacteria *E. coli*, which is one of the major pathogens associated with waterborne diseases [13]. The golden standard of their detection and quantification is by culturing the sample on a nutrient medium and counting the Colony Forming Units (CFU). Although the only reliable method, it is also the most demanding in terms of human resources, laboratories, and time required to obtain results, which is in the range of days.

Considering the previous facts, one can conclude there is a strong need for water quality monitoring to provide safety and security of water sources. This can only be achieved if there is a cost-effective system with a high spatial and temporal resolution and the possibility to monitor not only traditional water quality parameters such as turbidity, pH, and temperature, but also pathogen bacteria.

There are many commercial devices on the market designed to monitor water quality [6]–[11], [14]–[17]. However, their performance and/or prices represent a serious obstacle to the realization of real-time monitoring systems, especially those with high spatial and temporal resolution and the possibility to monitor the presence of bacteria. The devices [6]–[11] typically can measure 11-13 parameters, whereas only [6], [9] and detects the presence of bacteria. However, their prices start from €3,500, while the price range for most of them is from €7,000 to €10,000. Even in the case of the devices [14], [15] that measure 4 and 5 parameters, the prices are approximately €2,000 and €4,000, respectively. Moreover, all presented commercial solutions target in-situ water measurements conducted by humans with dedicated readers connected to the sensor. Although they can be used as a remote monitoring solution, they are not made with this concept in mind. They rather require additional gateways and communication equipment [18], [19] to make them Internet of Things (IoT) devices. Specifically, the Green Instruments G6200 water monitoring system [20] measures turbidity, pH, and temperature, but not the most critical parameter related to water pollution – the concentration of thermotolerant coliforms like *E. coli* bacteria. The Eureka Water Probes [21] instruments measure several water quality parameters. However, they are not autonomous. ECM ECO Monitoring a.s. [22] allows for remote transmission of the collected data, however, the data is related to only a few water quality parameters.

Scientific publications in recent decades report some progress, however, they are far away in terms of maturity from the commercial ones [18], [23]–[31]. Typically, they incorporate industrial sensors, a microcontroller unit, and wireless connectivity, while putting an accent on IoT and smart water quality monitoring systems. More concrete, they comprises off-the-shelf sensors for temperature [18], [23]–[31], pH [18], [23]–[30], turbidity [18], [23], [24], [27]–[30], and conductivity [18], [25], [26], [29], [31], while only some of the proposed solutions include sensors for oxidation reduction potential [18], [25], [31], dissolved oxygen [18], [26], [29], water level [23], [24], [29], flow rate [25], [27], phosphate [29], and total dissolved solids [30]. Although the presence of bacteria is a strong indicator of water quality, none of the papers addresses this problem. In terms of communication modules, most of the solutions rely on short-range WiFi including GSM communication technology [31], XBee [25], and Zigbee [29], whereas only sensor nodes presented in [24], [30] employ LoRa modules. Specifically, extension of LoRa coverage with LoRa repeaters with identification and selection of the best options for network elements spatial distribution is presented in [30]. A water quality monitoring in coastal areas is presented in [24]. The paper also provides a comparison of the LoRa network Received Signal Strength Indicator (RSSI) in the suburban coastal area from two different manufacturers of LoRa modules. Despite their great potential for better interpretation of the collected data, machine learning algorithms have been employed only in [28] and [30]. Finally, except for the solutions in [24], [30], none of the papers present field testing but only laboratory ones in controlled environments, which hinders the merit of those sensing nodes.

The aim of this paper is to present a novel, autonomous solution for real-time water quality monitoring that can measure four parameters, among which bacteria concentration, costs around 1500 EUR, employs machine learning approach for better data interpretation, provides in-field measurements, data collection and long-range remote transmission. Therefore, the solution addresses a strong demand for real-time water quality monitoring and previously described gaps existing in the state-of-the-art commercial and scientific solutions.

Specifically, this paper proposes a novel, economical, multiparameter water quality monitoring system for continuous monitoring of fresh waters. The system is based on a sensor node that integrates turbidity, temperature, and conductivity sensors, a miniature eighteen-channel spectrophotometer, and a sensor for the detection of thermotolerant coliforms. Having the ability to measure mud and algae concentration, temperature and total concentration of dissolved ions, the sensor node provides monitoring of standard water quality parameters. More importantly, the system contains a fluorescence sensor whose role is the detection of thermotolerant coliforms. Since the only direct method for bacteria detection requires laboratory facilities, the proposed sensor relies on an indirect method of Tryptophan-like fluorescence (TLF) [32]–[34]. It has been accepted as a good proxy for the detection of thermotolerant coliforms [35]–[37], although it is not an exact method for measurement of the number of bacteria.

The integrated sensors provide several different, yet not entirely uncorrelated signals, e.g., a higher concentration of mud and algae may give a false indication, or even cover the actual bacteria contamination. For that reason, a heuristic method of bacteria detection in the presence of mud and algae has been developed, which is another novelty of the proposed sensor node.

The sensor node is low-power and activates to measure and send data in periodic intervals while being in low-power mode in between. With an integrated Long Range Wide Area Network (LoRaWAN) module, we have realized a system that is wireless sensor network (WSN) ready and can send data to a dedicated server where the data is collected and further analyzed. In addition, the system can be submersed, and it is capable of long-term field operation.

The purpose of the system is to give early warning of incidental pollution situations. Thus, authorities can quickly respond by taking a water sample for laboratory analysis for confirmation, analyze the source of contamination, and take action regarding further prevention of such occasions. Featuring multiparameter detection including bacteria detection, heuristic analysis of the signals, long-term autonomous operation, wireless connection with a dedicated server, and an approximate commercial price of 1500 EUR, the proposed system overcomes the existing systems and present an excellent candidate for water monitoring systems with high spatial and time resolution.

The paper is organized as follows. After the Introduction, Section II provides an overview of the proposed system. A detailed description of the assembly of the sensor node is given in Section III, while the individual sensors are described in Section IV. Experimental results are shown in Section V followed by a discussion in Section VI. Section VII provides a conclusion. For the sake of clarity and a comprehensive explanation of the proposed system, the paper also includes two appendices dedicated to the calibration and temperature compensation of the sensors and the heuristic method.

## II. SYSTEM OVERVIEW

To develop a system that can address water quality monitoring in a quality and economical manner, a sufficient, yet not excessive number of water quality parameters should be chosen to be monitored. Also, the choice of the parameters depends on the type of monitored water, which, in our case, are fresh waters of high and good status. These types of water are usually clean; however, occasional incidents may occur which may contaminate them. With a low spatial and temporal resolution, such as manual sampling and especially long laboratory analysis, short-term pollution may not be detected, and properly addressed. These pollutions may come as excessive intake of water during heavy rains, melting of the snow, fecal contamination from the livestock, conscienceless actions of some individuals, etc. Therefore, it has been identified that the main pollutants to be monitored are fecal bacteria contamination (e.g. *E. coli*), algae, and mud.

In addition, the main requirements for the system are to be economically acceptable, to provide detection of required pollutants, to continuously, autonomously, and automatically monitor water quality, to be capable of remote access by IoT connectivity, to be submersible, and to be maintenance-free or to have long maintenance intervals. It is also important to state that the device is not meant to replace the laboratory analysis but to signalize potential pollution accidents.

Bearing in mind all the requirements, especially the price, we have developed a sensor node with five integrated sensors, all of which have been made in-house, i.e., no industrial or commercial sensors were used. In addition, the node is equipped with Arduino MKR WAN 1310 LoRa module, i.e., an integrated microcontroller module that has LoRa connectivity. Therefore, the device is capable of wireless sending of remote data to a LoRa gateway or to a LoRa network implemented within GSM network providers. The system is meant to be supplied from small local storage consisting of a battery and a renewable source such as a solar panel or a wind generator. To minimize power consumption, the system is operational only in predefined intervals, which is a configurable parameter, when measurements are taken and transmitted. The rest of the time the system is in sleep mode consuming minimum energy. A typical period when the system enters operational mode is around 1 hour. To meet the requirement to be submersible, the system has a buoy to maintain floating at a defined depth and strings for anchoring to the surrounding grounds.

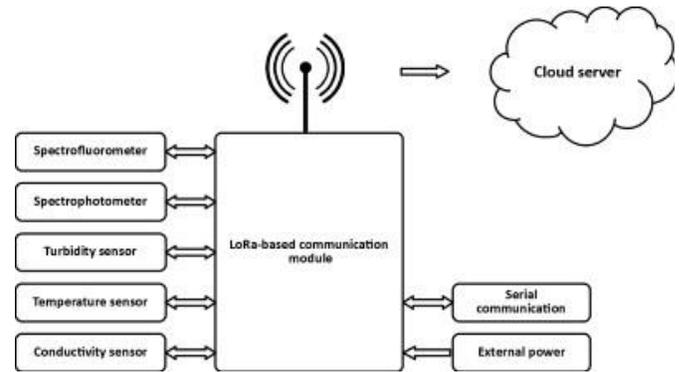

Fig. 1. Functional block diagram of the system.

The functional block diagram of the system is shown in Fig. 1, while Table I provides an overview of the sensors integrated into the sensor node. In the following sections, a detailed description of the design will be presented.

TABLE I
OVERVIEW OF THE INTEGRATED SENSORS

| Sensor | Measurement range | Detection level |
|---|---|---|
| Fluorescence sensor | 50 ppb of L Tryptophan | 6.25 ppb |
| Nephelometer | 500 NTU | 1 NTU |
| Spectrophotometer | 1000 a.u. (arbitrary unit) | 1 a.u. |
| Thermometer | -40-125°C | 0.0625°C |
| Conductivity sensor | 0-1000 µS/m | 1 µS/m |

## III. SENSOR NODE ASSEMBLY

To meet the requirements to integrate all individual sensors and supporting electronics in one sensor node that is to be submerged, a housing of the sensor node has been designed in a careful and cautious manner. The housing must safeguard the components, permit uninterrupted water flow through the measurement cuvette and prevent external light from entering the cuvette and potentially diminishing the accuracy of the measurements.

To safeguard the components after immersing the device up to one meter below the water's surface, a waterproof housing unit has been designed and fabricated using 3D printing technology. Fig. 2(a) and 2(b) show a model of the housing and a cross-sectional view of the essential housing parts respectively. The housing includes a central waterproof part that supports the measurement cuvette and electronics, and a waterproof lid. Both have been fabricated using SLA (Stereolithography) 3D printing technology and Formlabs Tough 2000 photo-resin material, characterized by excellent mechanical and chemical properties. The less demanding parts, such as the inlet and outlet, stabilizing fin, and hydrodynamic cover, have been fabricated using cheaper FDM (Fused Deposition Modeling) 3D printing technology and PETG (PolyEthylene Terephthalate Glycol) material, as waterproofing is not a requirement for these parts.

Fig. 3 shows an exploded view of all the components comprising the device (left), and a separate exploded view of the sensors and corresponding components used in and around the measuring cuvette (right). To support these components and ensure that all of them can be easily assembled and accommodated within the confined space of the central housing, a two-part fixture was designed and fabricated using FDM technology. Fig. 4 shows the central housing part (left) and fully assembled waterproof housing (right).

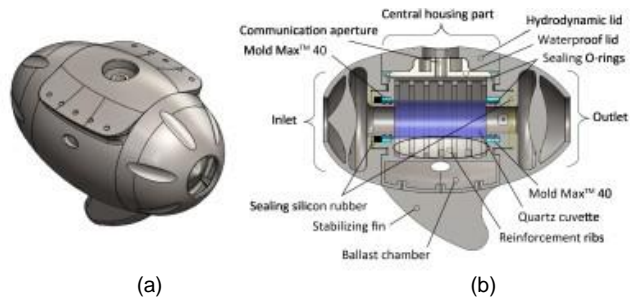

(a)                    (b)

Fig. 2. (a) Model of housing, (b) Cross-sectional.

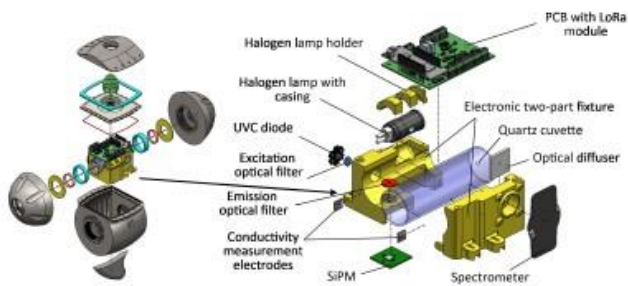

Fig. 3. Exploded view of all comprising components of the device. The left inset shows a cross-sectional view of the inlet, and the right inset shows exploded view of the two-part fixture with the accompanying assembled electronic components.

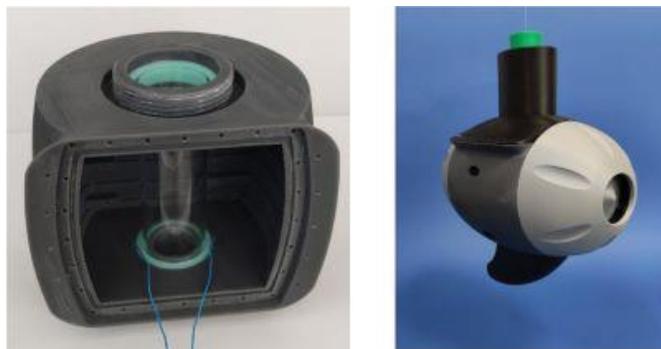

Fig. 4. Fabricated central housing part with installed measuring cuvette and electrodes for conductivity measurements (left) and fully assembled waterproof housing (right).

## IV. INDIVIDUAL SENSORS IN THE SENSOR NODE

The sensor node integrates five in-house, custom-made sensors, whose design will be shortly described here. Except for the thermometer, which is factory calibrated, all other sensors are calibrated using reference instruments and solutions. In the case of the fluorescence sensor, solutions of L-Tryptophan are used for calibration. A spectrofluorometer is used just to compare the responses since spectrofluorometers only provide arbitrary values of light intensity that also depends on the instrument's parameters setup. The integrated spectrophotometer is calibrated by adjusting the individual gains of all channels to obtain a flat response on all channels when deionized water was illuminated by the integrated halogen lamp. Detailed calibration procedures and temperature compensations are given in Appendix I.

### A. Fluorescence sensor

The major novelty of the proposed sensor node is the fluorescence sensor for the detection of the presence of thermotolerant coliforms, which is crucial to determine the quality of water. Since the direct method requires laboratory facilities, here we propose an indirect method based on measurements of the tryptophan-like fluorescence (TLF), which can be facilitated in a portable device. Namely, studies indicate that there is a fair correlation between TLF and the presence of fecal bacteria like *E. coli*, *Streptococcus,* and *Clostridium* [38], thermotolerant coliforms from ground waters [39], and biological oxygen from organic wastewaters [40]. While TLF is considered a good proxy for the detection, it is not an exact indicator. However, this is sufficient for this kind of device, which serves for monitoring and alarming of potential problems. In other words, the sensor is able to detect critical values of the concentration of the bacteria, and as such, is sufficient for the given purposes. Fig. 3 depicts the fluorescence sensor assembly.

In terms of design, the main issue of this method is low sensitivity, i.e., the ratio of excitation to emission signal, which is in the range of $10^6$ [41]. Therefore, it requires a very sensitive detector and careful design of the measuring cell. In such cases, a detector of choice is a high-sensitivity photomultiplier tube (PMT), which are very sensitive devices but they require high operational voltages (>1000V), [42]. Classical PMTs typically come as modules with an integrated power supply booster, however, they are bulky and expensive [43]. On the other hand, silicium-based photomultipliers (SiPM) have relatively low operation voltage, around 30 V, and a small footprint while maintaining the same properties as classical PMTs. Thus, ON Semiconductor MicroFC-60035 was chosen as the detector in the proposed fluorescence sensor.

As a light source, a UVC photodiode (UVR280-SA3P) with a central operating wavelength of 284 nm has been chosen and it is placed perpendicularly to the SiPM sensor. In that manner, a direct influence of the source light is minimized, and the sensor collects mostly fluorescence signals around 360 nm, but also to some extent a scattered light of 284 nm. Minimization of the influence of the source light is also achieved with two optical bandpass filters. An excitation filter is placed in front of the UVC diode, with a central wavelength of 280 nm, while the emission filter is placed in front of the SiPM sensor, with a central wavelength of 360 nm.

As the temperature of the system cannot be controlled, for the proper operation of SiPM temperature compensation is implemented, as detailed in Appendix I. The compensation is done in the electronics analog domain by varying the SiPM working voltage according to the temperature changes, which directly influences the amplification of SiPM. In addition, as SiPM exhibits self-heating during operation, the measurements are done fast with a long interval between measurements. During this interval, the generated heat during operation is evenly distributed to the SiPM surroundings enabling temperature compensation circuitry to work properly.

It is important to note that the detection of fluorescence coming from the presence of bacteria is sensitive to water purity. In particular, a higher amount of dirt and algae could give a deceptive impression of the presence of bacteria. To detect bacteria in the presence of mud and algae, a heuristic method has been created. The developed heuristic method is explained in detail in Appendix II.

### B. Spectrophotometer

Fig. 3 depicts the spectrophotometer construction. Its role in the proposed sensor node is the detection of dissolved minerals, in particular iron, manganese, and calcium ions [44]. A commercially available AS7265x multispectral chipset evaluation kit is used as a miniature discrete wavelength spectrophotometer. It consists of a set of photodetectors with eighteen channels, i.e., wavelengths. A miniature 5 W, 12 V halogen lamp (OSRAM 64405 S – 2700 K), which gives a continuous spectrum in the range 400-800 nm, is used to excite the spectrometer through a diffuser made of ALTUGLAS™ LED System Opal 3 mm sheet.

Although the number of channels may seem small when compared to laboratory spectrometers, during laboratory tests of water samples it has been found that the spectra of pollution materials that can be found in water, e.g., sludge, algae, humic acids, iron, and manganese salts have smooth spectra without distinct peaks, so sufficiently accurate results can be obtained with eighteen channels.

### C. Nephelometer

Nephelometer is used to measure the presence of suspended particles and it consists of a light source, and a light detector placed perpendicularly to each other. The halogen lamp, which also excites the spectrometer, is used as a light source, while an ambient light sensor ISL29023 is used as a photodetector. The sensor is placed at the bottom of the PCB and its infrared (IR) channel is used in this particular case.

### D. Thermometer

To reduce the complexity of the system, and with a fair assumption of temperature equilibrium, an integrated temperature sensor MCP9804 is implemented. The sensor is placed at the bottom of the PCB and is thermally coupled with the cuvette with a thermally conductive rubber sheet. In this manner, the water temperature is transferred to the cuvette and afterward to the sensor. As there are no abrupt changes in water temperature and the influence of self-heating is reduced with long inactive states between measurements it is expected that the temperature equilibrium is reached.

### E. Conductivity sensor

An in-house made conductivity sensor is based on two electrodes configuration and its role is to indicate potential pollution since an increased amount of dissolved solids entering waters usually results in increased conductivity. An AC square wave is used to excite the sensor and the detector measures the amplitude of the current going through the water between electrodes. The minimization of a double-layer capacitance effect and long-term stability is achieved with graphene electrodes.

We note here that except for the thermometer, all other sensors require calibration as they are custom-built with certain manufacturing differences. Besides, the spectrometer and the fluorescence sensor require temperature compensation. These aspects of the sensors are presented in detail in Appendix I.

## V. Device Testing, Validation, and Experimental Results

In this section, the laboratory test, validation, and field testing of the proposed sensor are presented. Prior to these the calibration and temperature compensation of each sensor was performed using commercial devices and reference laboratory instruments, as presented in Appendix I.

### A. Laboratory tests

The device was first tested in laboratory conditions on 55 water samples, that were taken from the water supply network of Ljubljana and Bled as well as from the localities Lipnik, Rečica Bled, Radovna, Sava Dolinka, Sava Bohinjka in Slovenia. Furthermore, we used the effluent water from the wastewater treatment plant in Domžale, 15 km northeast of Ljubljana, for testing purposes. We used two samples of concentrations of 250 ml effluent water in 2.5 l tap water and 1000 ml effluent water in 2.5 l tap water. The measurement results of all integrated sensors for the complete set of 55 samples are given in [45].

We note that the main aim of the experiments was to demonstrate the repeatability of the measurements, the robustness of the system, and measurement results that are in accordance with the incidental situations.

The repeatability of the measurement results can be seen in the example of the samples taken from the Bled water supply network, marked as Tap_Water_Bled 1-3. Fig. 5 presents responses of the spectrophotometer for the three samples and there is an excellent agreement of the spectrophotometer readings. Moreover, the repeatability is noticed in the case of the fluorescence sensor since the error relative to the full-scale output (FSO) is 0.09%. In the case of electrical conductivity the error relative to FSO is 0.14%, while the error relative to FSO of the nephelometer is below 0.05% [45].

To validate that the sensors change their responses with the change of samples' properties, samples of a mixture of wastewater with water from the city water supply network in Bled were analyzed. Wastewater was mixed with city water by adding 250 ml or 1000 ml of city water. Three repetitions of measurements were done for each sample prepared in this way. Similar repeatability can also be observed in the case of effluent samples, but for the sake of conciseness, we show only one sample of each case in Fig. 5.

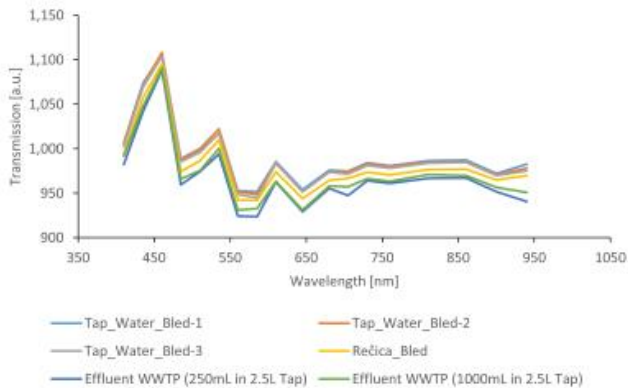

Fig. 5. Spectrophotometer readings for several different samples.

It is interesting to note that for the cases of Tap_Water_Bled-1, Effluent WWTP (250mL in 2.5 L Tap), and Effluent WWTP

TABLE II
FLUORESCENCE SENSOR, CONDUCTIVITY SENSOR, AND NEPHELOMETER MEASUREMENTS OF TAP AND EFFLUENT WATER MIXTURES

| Sample | Fluori meter [V] | Tryptophan [ppb] | El. conductivity [µS/cm] | Turbidity [NTU] |
|---|---|---|---|---|
| Tap_Water_Bled | 2.23 | 1.13 | 227.8 | 2.7 |
| Effluent WWTP (250mL in 2.5L Tap) | 1.50 | 16.93 | 509.1 | 3.8 |
| Effluent WWTP (1000mL in 2.5L Tap) | 0.01 | 49.81 | 620.8 | 4.8 |

(1000mL in 2.5 L Tap) fluorescence sensor responded by a decrease of the output voltage with increasing wastewater concentration, Table II. The table also contains the calculated values of tryptophan concentration based on the calibration curve presented in Fig. A.1(c), which demonstrates the possibility of detecting pollution that exceeds 6.9 ppb, a limit which water pollution is categorized as highly-critical (equivalent to >1 CFU/ml) [32]. The measured electrical conductivities show the expected trend of increasing values with increasing wastewater concentration, which is also noticeable with the turbidity sensor. It is important to note that turbidities below 4 NTU can only be detected by instruments, while at 4 NTU and above there are visible changes in water [46]. Therefore, in the case of the sample Effluent WWTP (250mL in 2.5L Tap), the water would seem drinkable, although the fluorimeter shows a higher concentration of tryptophan. There is a noticeable decrease in transmission on all spectrophotometer channels in the samples containing wastewater compared to the reference sample from the city network. This is a consequence of the increase in turbidity. The same stands for the electrical conductivity parameter which is permitted to be 2500 µS/cm. Both Effluent WWTP (250mL in 2.5L Tap) and Effluent WWTP (1000mL in 2.5L Tap) samples are within permitted limits, although tryptophan concentration is higher than allowed [46][45][45]. This illustrates the importance of monitoring tryptophan to ensure the water can safely be used for drinking.

The comparison of the sample Rečica_Bled-1 with the sample Effluent WWTP (250mL in 2.5L Tap)-1 is also interesting, since the surroundings of the location of the first sample are characterized by distinctly agricultural production, so the by-products of these activities end up in Rečica Bled. The spectra for both samples are shown in Fig. 5. The similarity between the samples is also evident when comparing the measurement results of fluorescence, electrical conductivity, and turbidity sensors [45].

*B. Validation*

We have done validation by analyzing and comparing the measurement results of the proposed system and the corresponding reference instruments. Two samples were taken from the Danube River in the city of Novi Sad, Serbia, whereas the first sample was taken from the muddy and sandy coast, while the second one was taken from the arranged part of the coast with a concrete riverbank.

The system was tested for the measurement accuracy of electrical conductivity, turbidity, and fluorescence. Reference measurements have been done using pH-EC-meter Hanna HI5522, turbidity meter WTW TURB 55IR, and spectrofluorometer FP-8558, respectively. Table III provides a comparison between the results obtained from the reference instruments and the proposed system.

TABLE III
FLUORESCENCE SENSOR, CONDUCTIVITY SENSOR, AND NEPHELOMETER MEASUREMENTS VERIFICATION RESULTS

| Sample | System/ Instrument | TLF [ppb] | TLF after postpro cessing [ppb] | El. conductivity [µS/cm] | Turbidity [NTU] |
|---|---|---|---|---|---|
| Sample 1 | Reference | 18.31 | 18.31 | 286.1 | 46.8 |
|  | Proposed | 13.98 | 14.74 | 286.5 | 46.85 |
| Sample 2 | Reference | 17.47 | 17.47 | 288.3 | 14.4 |
|  | Proposed | 14.95 | 15.05 | 288.2 | 14.35 |

Excellent agreement has been obtained for the electrical conductivity and turbidity, while a small deviation can be seen in the case of the fluorescence detection of Tryptophan-like fluorescence (TLF). By applying the heuristic method to the set of the device measurements the results of TLF become closer to the reference measurements, which is indicated as TLF after postprocessing in the table. Considering the discrete nature of bacteria, their low count, and non-uniform distribution in the samples, we can consider that these results are very well aligned with the referent ones.

To further elaborate on this matter, we have conducted microbiological analysis for total coliforms and *E. coli*. A volume of 1 ml from each sample was inoculated onto HiCrome™ Chromogenic Coliform Agar (CCA) (HiMedia Laboratories LLC, USA) plates. After a 24-hour incubation period at 37°C, as a chromogenic substrate is used, dark blue colonies were identified and counted, indicating the presence of *E. coli*. The red colonies represent non-categorized total coliforms, Table IV.

TABLE IV
RESULTS OF MICROBIOLOGICAL ANALYSIS FOR ESCHERICHIA COLI AND TOTAL COLIFORMS

| Sample | E. coli [CFU/ml] | Total coliforms [CFU/ml] |
|---|---|---|
| Sample 1 | 1 | 85 |
| Sample 2 | 2 | 47 |

It can be observed that the detected number of *E. coli* in 1 ml is similar in both samples. Following the classification presented in [48], the samples fall into Classes I and II which are below the fecal pollution threshold, maintaining suitable quality for bathing water (*E. coli* CFU < 10 per 1 ml), which is aligned with the standards outlined in the EU directive [49].

While there is no straightforward relation between the units ppb and CFU/ml, this additional experiment can be used to confirm that the proposed device has correctly indicated the presence of the *E. coli*. Namely, a critical concentration value above which water pollution is categorized as highly-critical equals 6.9 ppb, which is roughly equivalent to 1 CFU/ml [32]. The values from the tables show that the device correctly indicates the pollution, however, with slightly higher values of TFL than expected. This slight disagreement can be attributed to the presence of the other unknown coliforms in the samples since we cannot eliminate the possibility of their influence on the measurements of the TLF. One should note that, however, those coliforms can also be considered as a health hazard.

### C. Field tests

In addition to the laboratory measurements and validation, field measurements were carried out. Initial testing of the device in real conditions was planned on site Ribiška Družina Bled-1, Slovenia (46°22'14.4"N 14°05'15.1"E). The device was connected to a battery power supply that is charged by means of a solar panel. The device was placed in the metal frame and fixed with plastic ties, and as such submerged in the water, Fig. 6. The power supply and antenna cable were passed through the green plastic pipe. The data were collected in real-time every 5 minutes and sent to the MILESIGHT LoRaWAN Gateway UG65-868M-EA, which is connected to the ZTE MF971R LTE modem. The modem forwarded data to the server via the Telekom Slovenije internet connection.

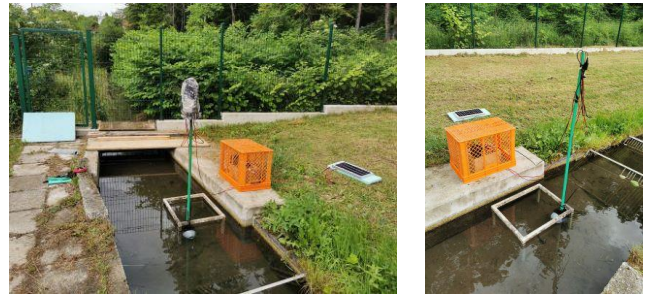

Fig. 6. Setup of the device at the testing site located in Ribiška Družina Bled.

The device was first tested for a period of twenty hours, during which the increase of water turbidity was artificially induced by its manual mixing upstream of the device. This moment corresponds to the date 06/09/2022 and time 18:37, which is clearly indicated in Fig. 7(c) showing a sudden jump to 41.6 NTU.

After this initial test with artificially induced turbidity, a sludge deposition process occurred, which can be seen by the gradual increase of the turbidity in Fig. 7(c). This is also reflected in the measurement results of the fluorescence sensor where the voltage values drifted until the voltage output of the sensor settled at 2.45 V, which corresponds to the situation when there is no fluorescence, i.e., when the sensor is optically blocked. The sludge deposition was additionally promoted by the cleaning of the surrounding channels and the mowing of the nearby grass, which caused another jump in the turbidity to 163.9 NTU. This happened on 10.06.2022 at 12:52, Fig. 7(c). This intentionally caused turbidity and the consequences of channel arranging, led to the sludge deposition at the bottom of the cuvette, where the fluorescence sensor is located, blocking the optical path to it. This outcome was expected because the device was designed for the detection of excessive situations, after which it is necessary to clean it and then return to the desired location to be active again. In order to confirm this statement, the device was cleaned, and afterward, the fluorescence sensor measurement returned to the default values.

Finally, long-term testing was carried out in the form of continuous measurement in the period from June 17 to July 31, 2022, with a frequency of 5 minutes. Overall results can be found in [45]. The purpose of this test was to inspect the durability of the device as well as changes in measurement results caused by changing weather conditions. The location was the same as in the previous case, Ribiška Družina Bled. The readings of the spectrophotometer, fluorescence sensor, turbidity sensor, and electrical conductivity sensor for the specified period are shown in Fig. 8, respectively.

According to the results of the spectrophotometer, there are certain situations where a sudden drop in transmission happens, and these situations are associated with the jumps in turbidity. The ripples that occur in the transmission, which are most pronounced for the 460 nm channel, are the result of water temperature fluctuations. It is important to note that this phenomenon would have been even more pronounced if the temperature compensation of the device was not done.

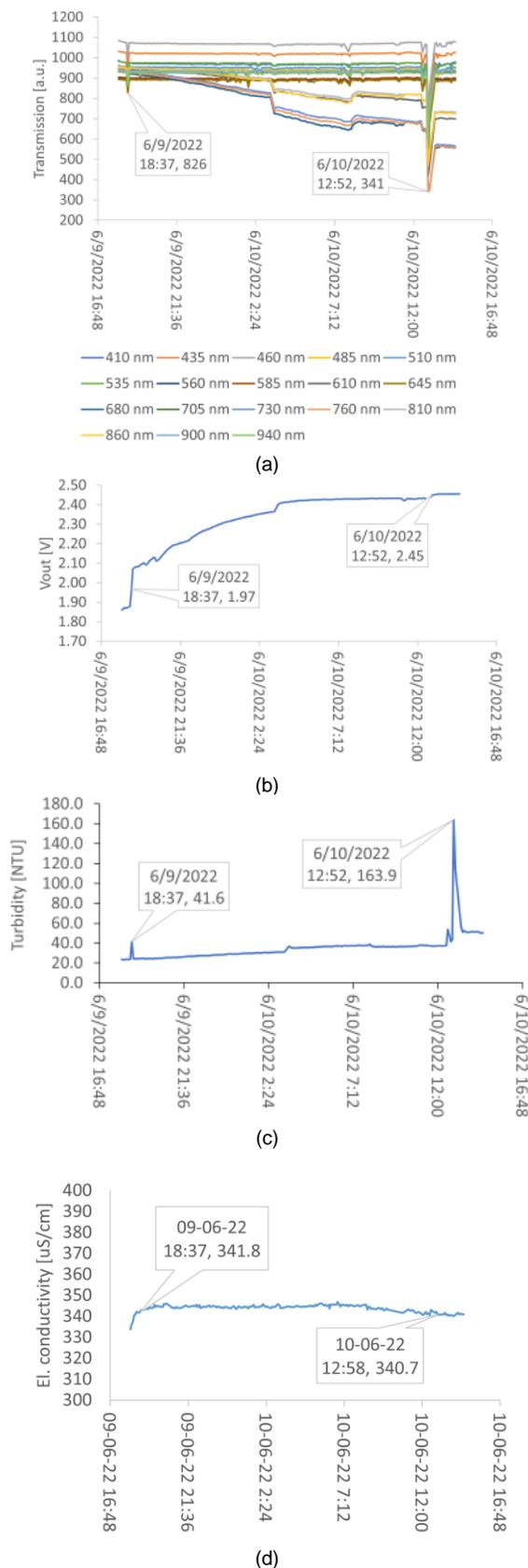

Fig. 7. Initial test on the device done on site Ribiška Družina Bled: (a) Spectrophotometer, (b) Fluorescence sensor, (c) Nephelometer, (d) Electrical conductivity sensor.

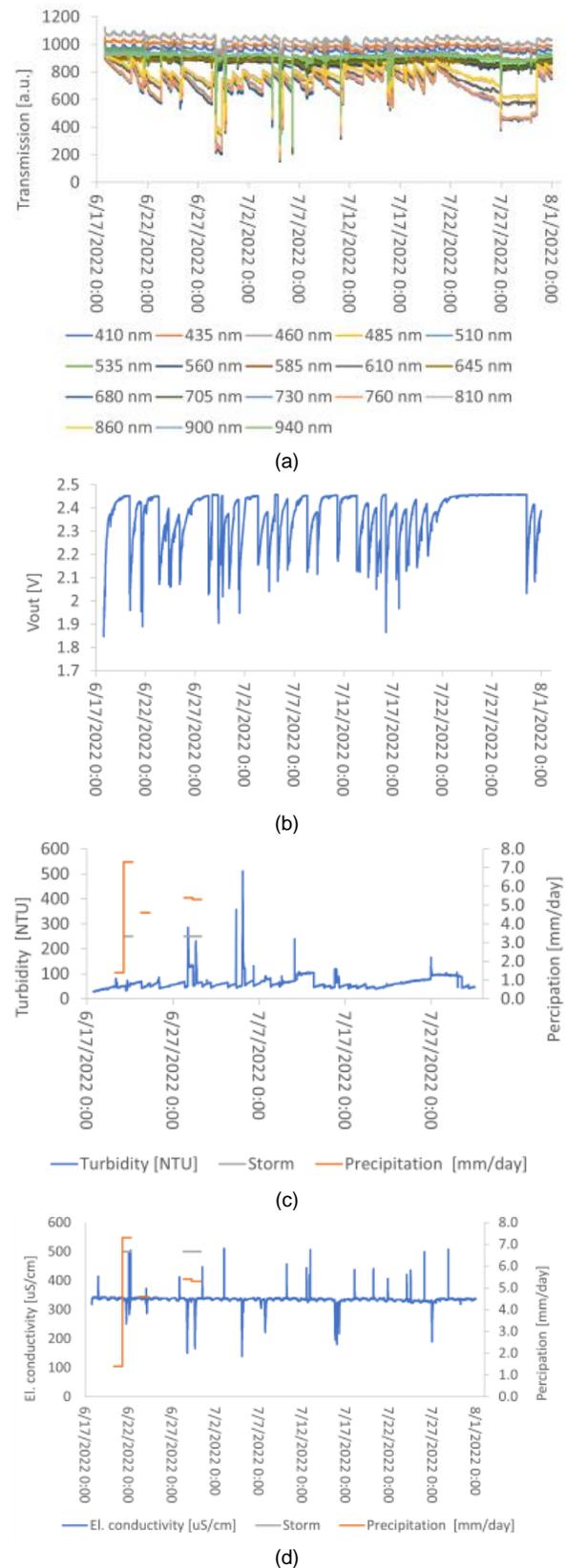

Fig. 8. Measurements at the Ribiška Družina Bled site between June 17 and July 31, 2022: (a) Spectrophotometer, (b) Fluorescence sensor, (c) Nephelometer, (d) Electrical conductivity sensor.

Using the data from the weather station of the Environmental Protection Agency of the Republic of Slovenia (ARSO – Agency of the Republic of Slovenia for the Environment), the comparison of the amount of precipitation and storm events with the data shown by the sensors on our device was done. Fig. 8(c) shows a comparison with the nephelometer in the period from June 17 to July 3, 2022. It can be clearly seen that the occurrence of precipitation with a certain time delay causes an increase in turbidity, which is especially pronounced in the presence of a storm. There is a time delay in the readings of the nephelometer because a certain period is needed from the moment it starts raining to the moment it starts pouring and muddying the water. Also, it can be seen that precipitation causes a decrease in the value of electrical conductivity, Fig. 8(d).

It can be concluded that the complete system is reliable and robust. Due to the measurement frequency of 5 minutes, the system was able to record even short-term periods of water pollution that lasted less than 15 minutes. Data from ARSO meteorological stations show a clear influence of weather conditions, which influenced sensors' readout.

## VI. Discussion

The sensor node presents a novel, economical, multiparameter water quality monitoring solution that can provide low-power, autonomous operation, wireless data collection and remote transmission. The node integrates turbidity, temperature, and conductivity sensors, a miniature eighteen-channel spectrophotometer, as well as a sensor for the detection of thermotolerant coliforms, which is the major novelty of the node. Due to the sensors' signals correlation, a heuristic model has been developed to provide an accurate interpretation of the sensors' readings, which is another novelty of the proposed system. Moreover, the estimated price of the developed prototype is around 1500 EUR.

Experimental results have confirmed the robustness and reliability of the system as well as the expected behavior when it comes to the changes in the analyzed water sample. More concretely, the increased contamination by effluent water or artificially induced increase of water turbidity caused the expected changes in the spectrophotometer readings as well as in fluorescence, conductivity, and turbidity sensors' measurements. The validation results and comparison to the reference instruments indicate the accuracy of the device. There is only a slight disagreement in the TLF measurements, which can be attributed to the presence of various coliforms.

We would like to note that every outdoor system is exposed to varying environmental conditions that can affect the system's performance. Depending on the system, different environmental conditions can have an effect, however, the common one for all is temperature. With the temperature compensation of all measurements within the extended temperature range (4-40°C), the temperature influence in this system has been minimized.

In comparison to the solutions presented in the scientific publications, the main advantage of our solution is the detection of *E. coli*, which is a parameter of crucial importance for water quality monitoring. Together with [24], [30], the proposed device is the only one that employs LoRa modules, which have longer range in comparison to WiFi modules and provide lower power consumption. Also, the proposed solution is the only one that has provided not only laboratory tests but also long-term in-field operation. In addition, a machine learning algorithm is used to provide better data interpretation, which was only used in [28] and [30]

Compared to the commercial solutions [6]–[8], [10], [11], [14]–[17] our system provides integrated wireless connectivity and total coliform/bacteria detection, which is a crucial human safety parameter. At the same time, it is expected that the commercial sensor node will cost around 1500 EUR, which is significantly lower than the costs of the solutions available on the market. Depending on the configuration, the number of parameters of the commercial solutions can be higher compared to our solution, however, the commercial price can be significant. The intention of this work is not to completely compete with the commercial solutions, but to provide sufficiently accurate, affordable, remote sensing solution for early detection of incidental water contamination.

With respect to total coliform/bacteria detection, to our knowledge, there are only two commercial instruments for instantaneous in-situ measurements, [6] and [9]. The Proteus probe has double the range compared to the proposed system, however, this range can be extended/adopted in our system by changing the gain. We chose the lower range since the proposed system is targeting clean and fresh waters, and the limit for highly contaminated water according to WHO is set to only 6.9 ppb. Both implement temperature compensation, however, the proposed system utilizes analog compensation, whereas the Proteus probe utilizes digital compensation. Correction of the measurements due to the water impurities is utilized with a turbidity sensor in the Proteus probe, whereas the proposed system, besides turbidity, utilizes signal from the integrated spectrophotometer which takes algae measurements into account. Finally, the expected commercial price is significantly lower than that of [9].

As for the solution [6], there is limited information about its performance. It can measure more parameters and its bacteria concentration range is greater than those of the proposed one. However, this range can be extended in the proposed solution, and the expected commercial price of the proposed solution is 6 times lower than that of [6].

Although the entire system was successfully tested in an operational environment, based on the tests performed, it can be concluded that there is a possibility to further improve the performance of the system. Namely, the fluorescence sensor showed sensitivity to sludge deposits mainly due to the physical position of the sensor at the bottom of the cuvette. It is important to note that the chosen location of Ribiška Družina Bled was extreme for this device, because the water contained a significant amount of silt, while the water flow was extremely small. These are suitable conditions for sludge deposition; therefore it was necessary to shake the device daily while it is in the water to promote the flow of water and clean the cuvette. At this location, for the observed period, the turbidity sensor showed an average value of 65 NTU, which is a significantly higher value of turbidity than the range of up to 10 NTU for which the device was designed. Certainly, these are not the conditions expected of the first-class mountain rivers for which this device is intended. If these conditions were matched, the necessary maintenance period would be on a monthly basis.

To further improve the robustness of the system and reduce the need for regular maintenance, it is necessary to either rotate the sensor configuration by an angle of 45° around an axis located along the central axis of the cuvette or place the fluorescence sensor on the upper side of the cuvette. This would require additional changes in the printed circuit board of the accompanying electronics. The mentioned changes may be the subject of future activities aimed at a more advanced version of the prototype.

Furthermore, our future work will focus on comprehensive field testing of the device across different environmental conditions and water bodies to validate its accuracy and reliability. To make the system more user-friendly and useful for a wide range of potential users, development of automated data analysis, alert issuing, user-friendly interface and mobile application will be considered. Considering data remote transmission and sensitivity of the collected data, we will also focus on development of the protocols to ensure data security, data integrity and confidentiality.

## VII. Conclusion

We have presented a novel, economical, reliable, and low-power water quality monitoring system that can operate autonomously and provide wireless data collection as well as high spatial and time resolution. Besides mud and algae concentration, temperature, and total concentration of dissolved ions, the sensor node provides detection of thermotolerant coliforms, with the accuracy that is sufficient to alarm incidental pollution situations in waters of class I and II. The integrated sensors provide several distinct, yet not entirely uncorrelated signals, and thus, a heuristic method of bacteria detection in the presence of mud and algae has been developed, which is another novelty of the proposed sensor node.

A detailed description of the overall system, individual sensors, their calibration, and temperature compensation has been provided together with a detailed explanation of the heuristic method. The experimental validation of the sensor node has been conducted at a number of localities in Slovenia and one in Serbia. Those experiments have confirmed the robustness, reliability, and stability of the system. The measurement results of the spectrophotometer, fluorescence, conductivity, and turbidity sensors have exhibited behavior that was in accordance with the changes in the analyzed water samples. The results related to the samples taken in Serbia and comparison to the reference instruments indicate the accuracy of the device. There is only a slight disagreement in the TLF measurements, which can be attributed to the presence of various coliforms.

Although there is room for improvement, the system has clearly demonstrated the ability to be used for early warning of incidental pollution situations. In such scenarios, it would enable authorities to fast respond by taking a water sample for laboratory analysis for confirmation, analyze the source of contamination, and act regarding further prevention of such occasions.

## Appendix I - Calibration and Temperature Compensation

Except for the thermometer, all sensors require calibration as they are custom-built with certain manufacturing differences. The conductivity sensor has been calibrated at room temperature. The reference equipment for the electrical conductivity used for calibration was Hanna HI5522. Salt (HCl) solutions, with geometric progression with a common factor of 2, were prepared and used for calibration. Fig. A.1(a) shows a calibration curve for one of the systems as well as calibration coefficients for the second-order calibration curve. Due to rectification of the square wave signal an offset of half measurement range is introduced.

The nephelometer has also been calibrated at room temperature with the help of the referent instrument (turbidity meter WTW TURB 55IR), and geometric progression values of turbidity for calibration have been used, Fig. A.1(b).

Besides calibration, the spectrometer and the fluorescence sensor require temperature compensation. To that end, an environmental chamber and corresponding temperature profile have been used, Fig. A.2. The temperature profile covers and extends even beyond expected water temperatures. The temperature gradient is adjusted such that the system temperature can follow the set temperature, and the used samples of water and solutions of interest are also kept in the chamber.

The fluorescence sensor is temperature compensated in the analog part of the system. Based on calculation with respect to gain and offset temperature change of SiPM an initial compensation is made. Afterwards, a fine-tuning is done in a few iterations in the environment chamber, which can be considered to have sufficient accuracy. Table A.I shows measurements of three different concentrations of tryptophan at three different temperatures. Fig. A.1(c) shows the calibration curve for the fluorescence sensor where the calibration values are taken as an average sensor readout at three different temperatures. One should note that the negative slope stems from the realization of analog circuitry.

TABLE A.I
MEASURED VALUES OF FLUORESCENCE WITH THREE DIFFERENT CONCENTRATIONS AT THREE TEMPERATURES

| Tryptophan [ppb] | ADC value @5°C | ADC value @10°C | ADC value @15°C |
|---|---|---|---|
| 0 | 2.26885 | 2.27804 | 2.26306 |
| 15 | 1.52333 | 1.56176 | 1.66162 |
| 30 | 0.73572 | 0.827642.26306 | 0.94348 |

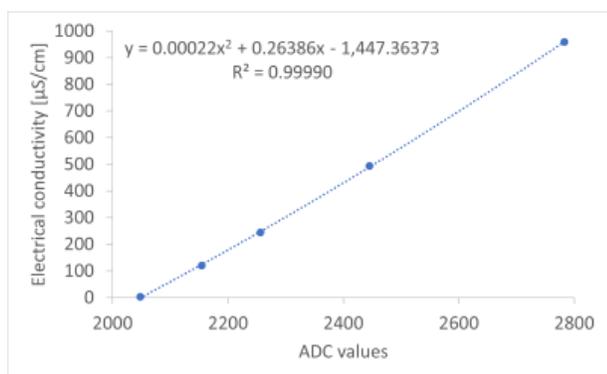

(a)

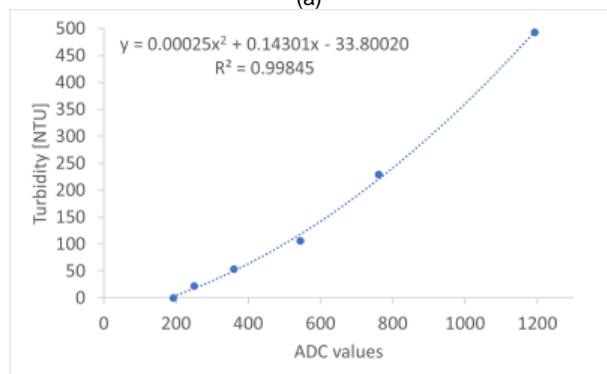

(b)

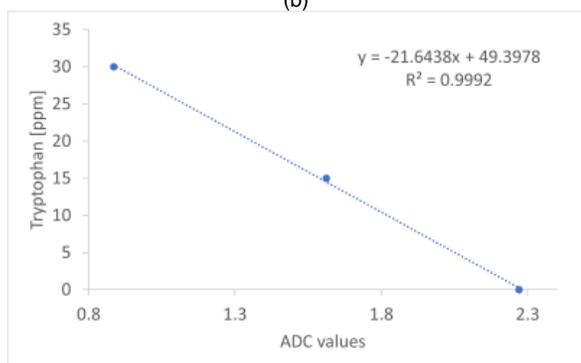

(c)

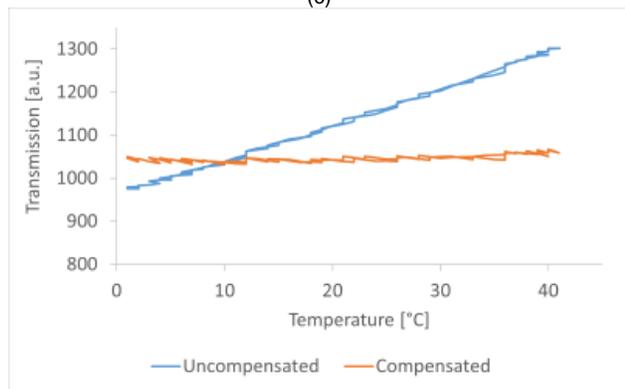

(d)

Fig. A.1. Calibration curves of: (a) Conductivity, (b) Nephelometer, (c) Fluorescence, (d) Spectrophotometer sensors.

To determine the quality of the fluorescence measurement of the device, a comparison was made with a reference instrument (spectrofluorometer FP-8558, company JASCO). All tested samples were made in deionized water without any admixture of sludge, algae, or other impurities, except for L-Tryptophan, since the reference device does not have any correction for these impurities. Comparative measurements were made on samples with different concentrations of L-Tryptophan (0, 6.25, 12.5, 25 and 50 ppb). Since in both cases it is an optical method, where the intensity of light is measured, the outputs of both devices are dimensionless. Therefore, both sets of measurements were normalized to the range from 0 to 1, taking the response to zero concentration as a reference. Fig. A.3 shows the results of comparative measurements with the reference device and the developed device. The matching of the results is excellent, moreover, the developed device shows better linearity ($R^2$ = 0.9998) compared to the referent one ($R^2$ = 0.9964).

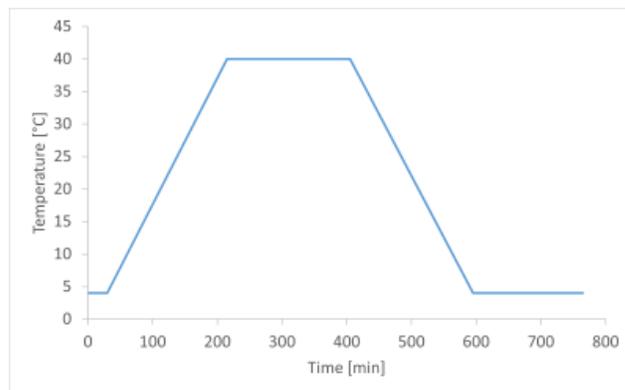

Fig. A.2. Environmental chamber temperature profile

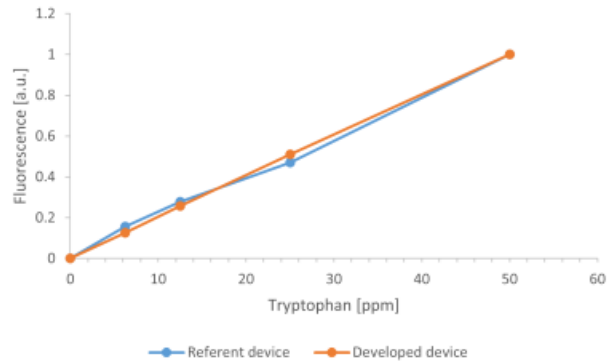

Fig. A.3. Comparative fluorescence measurements for different tryptophan concentrations with referent spectrofluorometer and the developed device.

The spectrophotometer is calibrated by setting internal gain coefficients of all channels. This is done at the temperature of 10 °C and the gain coefficients are calculated such that the reading of deionized water gives 1000 counts or arbitrary units. Fig. A.1(d) shows readings of one channel over the temperature range 0-35 °C, whereas the channel is calibrated at 5°C. The readings show a high temperature dependency of nearly 25% in the given temperature range. However, concerning the linearity of the temperature dependency, digital compensation is easily achieved. Fig. A.1(d) also shows the same readings after the digital compensation, which provides almost a flat curve.

The electrical conductivity of water is temperature dependent and described in the literature [50]. Therefore, the system provides two values of electrical conductivity, the raw value based on calibration, and EC25, or calculated electrical conductivity at 25°C based on the raw value and temperature measurements.

## APPENDIX II – HEURISTIC METHOD

The detection of fluorescence coming from the presence of bacteria is sensitive to water purity. A higher concentration of mud and algae may give a false indication or even cover the actual bacteria contamination. For that reason, a heuristic method of bacteria detection in the presence of mud and algae has been developed.

Once the calibration process of integrated sensors is finished as described in Appendix I, an experiment that simulates actual environmental conditions is conducted. We created a dataset containing sensor measurements of prepared water samples with three different concentrations of L-tryptophan, as the model for fecal bacteria [21]–[23], and algae, and mud, as contaminants that negatively influence the measurement process. Selected concentrations of L-tryptophan are in the range of [0, 30 ppb], which includes a critical concentration value of 6.9 ppb above which water pollution is categorized as highly-critical (equivalent to >1 CFU/ml) [32]. The concentrations of algae and mud are selected within the following ranges [0, 200 kj/ml] and [0, 100 mg/l], respectively.

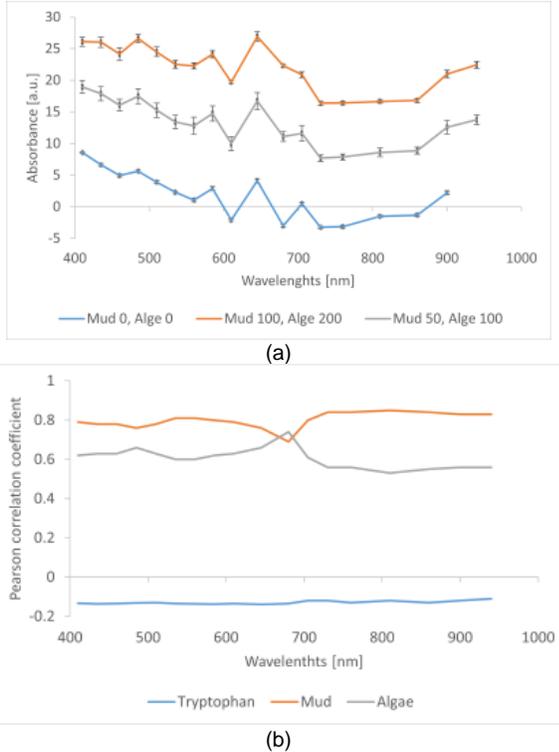

Fig. A.4. (a) Responses of the spectrophotometer (absorption) for mean and limit values of algae and mud concentration, at all possible values of L-tryptophan concentration, (b) Pearson's correlation coefficient between spectrophotometer absorption at different wavelengths and concentrations of L-tryptophan, algae, or mud.

Fig. A.4(a) shows the mean values of spectrophotometer response with standard deviation for all selected eighteen channels. The responses were calculated for selected combinations of algae and mud concentration and for all combinations of L-tryptophan concentration. The largest change in the absorption signal trend is present between the wavelengths of 680 nm and 705 nm, with a tendency for the absorption signal to increase as the concentration of algae and mud increases. Fig. A.4(b) shows a Pearson's correlation coefficient between values of absorption for each channel, and selected concentrations of L-tryptophan, algae, and mud. Low correlations between L-tryptophan concentration and spectrophotometer responses indicate that they are not reliable predictors of L-tryptophan concentration. On the other hand, the variability in the correlation coefficient between the concentration of mud or algae and the absorption at all channels of the spectrophotometer, and its high values (for some channels > 0.8), indicate that the presence of algae and mud can be reliably predicted in water samples based on spectrophotometer responses.

Therefore, we propose a method for estimating the concentration of L-tryptophan in two steps. The first one is the development of models for prediction of algae and mud concentration based on responses of spectrophotometer. The second step is related to the creation of a model for correcting the response of fluorescence sensor based on the estimated concentrations of algae and mud.

Within the first step we adopt a polynomial regression model for prediction of the concentration of mud and algae. Based on the graphs in Fig. A.4, the absorptions at 560, 680, and 810 nm show opposite correlations with the concentrations of mud and algae. This contributes to the discrimination between the appearance of mud and algae. First, we introduce the basic set of features within vector $s \in R^k$, ($k = 5$) consisting of absorption values of spectrophotometer at 560 nm, 680 nm and 810 nm ($s_{560}, s_{680}, s_{810},$) and responses of nephelometer $s_{NF}$ and conductivity sensor $s_{CS}$

$$\boldsymbol{s} = [s_{560}, s_{680}, s_{810}, s_{NF}, s_{CS}]^T. \quad (1)$$

The basic features from $\boldsymbol{s}$ are further extended by including bias term and introducing polynomial combinations with degree $\leq 2$ of features related to spectrophotometer responses in $\boldsymbol{s}$. We obtain the final feature vector $\boldsymbol{v} \in R^d$, ($d = 12$):

$$\boldsymbol{v} = [1, s_{560}, s_{680}, s_{810}, s_{560} * s_{680}, s_{560} * s_{810}, s_{680} * s_{810}, s_{560}^2, s_{680}^2, s_{810}^2, s_{NF}, s_{CS}]^T, \quad (2)$$

comprising components of $\boldsymbol{s}$, square of each component in $\boldsymbol{s}$ and intermediate products between every two of them. The polynomial regression coefficients are estimated using the ordinary least squares (OLS) method with L2 regularization. Since we have a concentration of algae and mud denoted as $c_a$ and $c_m$, respectively, as two target variables, coefficients for both polynomial regression models are stacked as columns of a matrix $\boldsymbol{A} \in R^{d \times 2}$. This is further used for estimating concentrations of algae and mud as:

$$\begin{bmatrix} \hat{c}_a \\ \hat{c}_m \end{bmatrix} = \hat{\boldsymbol{A}}^T \cdot \boldsymbol{v}, \quad (3)$$

where $\hat{\boldsymbol{A}}$ represents estimated regression coefficients for both models independently and $\hat{c}_a$ and $\hat{c}_m$ are estimated concentrations of algae and mud, respectively. Finally, we propose a linear regression model as a second step of our method, for correcting the response of fluorescence sensor denoted as $r_{fl}$ based on estimated $\hat{c}_a$ and $\hat{c}_m$ as follows:

$$\hat{c}_{TLF} = \beta_0 + \beta_1 \cdot r_{fl} + \beta_2 \cdot \hat{c}_a + \beta_3 \cdot \hat{c}_m$$
$$= f(r_{fl}, \hat{c}_a, \hat{c}_m), \quad (4)$$

and thus, estimating concentration of L-tryptophan denoted as $\hat{c}_{TLF}$. Regression coefficients $\beta_{0-3}$ of the model $f(r_{fl}, \hat{c}_a, \hat{c}_m)$ are estimated using OLS method with L2 regularization.

Using cross-validation with a leave-one-out test sample from the formed dataset, the polynomial regression coefficients are estimated and stacked as columns of matrix $\hat{A}$. The standard deviation of the estimated coefficients does not exceed the value of 0.02. Based on the obtained mean absolute values of the estimated coefficients after cross-validation, the responses of the conductivity sensor and the nephelometer have the greatest influence on the estimation of the concentration of algae and mud. In addition to the response of the conductivity sensor, the assessment of algae concentration is most influenced by the absorption response of the spectrophotometer at 680 nm, i.e. created features $s_{680}^2$ and $s_{680}$. The prediction error of algae and mud concentration based on the estimated matrix $\hat{A}$ and the prepared feature vector $v$ after the experiments with the omission of one sample for the test is $0.16 \pm 0.1$.

During the experiment for dataset formation, the responses of the fluorimeter $r_{fl}$ for different combinations of concentrations of L-Tryptophan $c_{TLF}$, algae $c_a$ and mud $c_m$ are recorded.

Knowing the actual values of $c_{TLF}$ in prepared laboratory water samples, at selected different combinations of $c_a$ and $c_m$ ratios, and the $r_{fl}$ for the observed samples, the correction factor for the measured values of the fluorimeter is determined through the linear regression model $f(r_{fl}, \hat{c}_a, \hat{c}_m)$. With the estimated matrix $\hat{A}$, estimated concentration of algae and mud $(\hat{c}_a, \hat{c}_m)$ are obtained using (3), and then regression coefficient $\beta_{0-3}$ for $f$ are estimated. Table A.II gives mean values and standard deviations of estimated $\beta_{0-3}$ after cross-validation with leave one-out test sample on formed dataset.

TABLE A.II
THE FIRST ROW: INPUT FEATURES TO $f$, USED TO ESTIMATE THE CONCENTRATION OF L-TRYPTOPHAN $c_{TLF}$. THE SECOND ROW: STATISTICS OF LINEAR REGRESSION COEFFICIENT VALUES $\beta_{0-3}$ OF MODEL $f$.

| Inputs for $f$ | $r_{fl}$ | $\hat{c}_a$ | $\hat{c}_m$ | bias |
|---|---|---|---|---|
| $\beta_{0-3}$ | $\beta_1 \sim (-0.68 \pm 0.02)$ | $\beta_2 \sim (-0.22 \pm 0.01)$ | $\beta_3 \sim (-0.06 \pm 0.01)$ | $\beta_0 \sim (0.78 \pm 0.02)$ |

The total mean square error of L-Tryptophan concentration estimation from several experiments with the omission of one test example is $0.022 \pm 6.35\mathrm{e}^{-18}$. The estimated error is within 2.5% of the maximum tryptophan concentration used in the experiments with negligible variation.

## ACKNOWLEDGMENT

We would like to thank Dr Ljiljana Janjusevic for the help in microbiological analysis for total coliforms and *E. coli*.

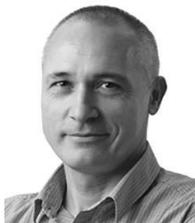

**Damir B. Krklješ** received a Ph.D. degree from the University of Novi Sad, Faculty of Technical Sciences, Novi Sad, Novi Sad, Serbia, in 2016.

He worked as a Research and Teaching assistant at the Faculty of Technical Sciences, teaching electronics, sensors and actuators, and mechatronics. Starting in 2013 he moved to the industry, where he worked for a few companies, mostly in a design service area for leading worldwide industrial partners.

He currently works as a Research Associate in the Centre for Sensing Technologies at the BioSense Institute, Novi Sad, Serbia. He is a coordinator for Robotics and Applied Electronics in agriculture research fields. He published over 30 journal and conference papers.

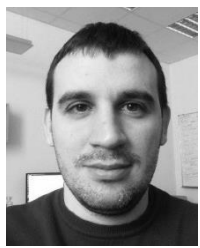

**Goran V. Kitić** received a Ph.D. degree from the Faculty of Technical Sciences, University of Novi Sad, Novi Sad, Serbia in 2016 on the subject of microwave soil moisture sensors.

He currently works as a Senior Research Associate and Head of the Center for Sensing Technologies at the BioSense Institute, Novi Sad, Serbia. He developed experience in the field of different fabrication technologies and characterization methods. His research interests are related to the development of sensors applied in agriculture, the food sector, and environmental protection. Some of the developed sensor prototypes include a multispectral sensor for plant status assessment, a soil moisture sensor, nitrates in water detector, and agricultural robots. His research work resulted in eight international journal publications, and numerous papers published at international conferences.

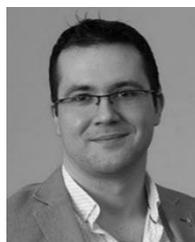

**Csaba M. Petes** received B.S. in 2014 and M.S. degree in 2017 from the Faculty of Technical Sciences, Department of microprocessor systems and algorithms, University of Novi Sad, Novi Sad, Serbia.

He currently works in the Centre for Sensing Technologies at the BioSense Institute, Novi Sad, Serbia. His research is related to embedded software and robotics systems, where he participated in many projects. In those projects, he primarily developed analog and digital electronics for read-out systems, designed PCB layouts, realized PCB prototypes, and supported the overall design of robots, integration, and signal processing.

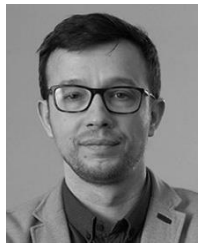

**Slobodan S. Birgermajer** received a Ph.D. degree from the Faculty of Technical Sciences, University of Novi Sad, Novi Sad, Serbia in 2018.

He currently works as a Research Associate in the Centre for Sensing Technologies at the BioSense Institute, Novi Sad, Serbia. He authored and co-authored 11 international journal papers, 11 international conference papers, 11 technical solutions, and 1 domestic patent application. He participates in different national and international projects (H2020, FP7, RISE, Eureka). He is currently working on the development and applications of sensors in agriculture, the food industry, and environmental monitoring, His main research interest includes microwave passive devices, photonic crystals, microfluidic sensors, 3D and 4D printing technologies, and fiber optics.

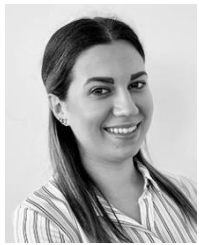

**Jovana D. Stanojev** received a Ph.D. degree from the Department of Materials Engineering, Faculty of Technology, University of Novi Sad, Novi Sad, Serbia in 2020.

Her current research focuses are the development of new materials for sensors, nanomaterials and nanomembranes, microfluidics, the development of organ-on-a-chip platforms, and cell culturing. She has 10 scientific papers and numerous conference papers, with an h-index of 4 and about 70 citations (Google Scholar).

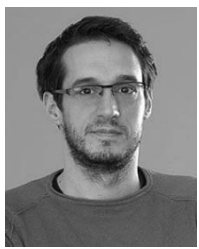

**Branimir M. Bajac** received a Ph.D. degree from the Faculty of Technical Sciences, University of Novi Sad, Novi Sad, Serbia in 2017 in the field of materials engineering.

He currently works as a Research Associate in the Centre for Sensing Technologies at the BioSense Institute, Novi Sad, Serbia. His primary research focus is on the deposition and characterization of solid inorganic thin films of single and mixed oxides. He possesses a strong knowledge of sensor design and an overview of state-of-the-art sensors applied in agri-food chains. His current interest is in the field is in environmental sensors design and application of such in IoT networks for continuous monitoring of water and air quality.

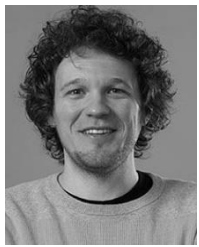

**Marko N. Panić** received a joint Ph.D. degree from the Faculty of Technical Sciences, University of Novi Sad, Novi Sad, Serbia and Ghent University, Ghent, Belgium in 2020.

He currently works as a Research Fellow in the Centre for Information Technologies at the BioSense Institute, Novi Sad, Serbia. He works on advanced statistical analysis of image data, obtained from different sensors, using machine learning techniques with applications in biology, agriculture, environmental sciences, and health.

Dr. Marko Panić, as a member of the BioSense team, participated in the Syngenta Crop Challenge three times, which resulted in winning first and third place and two OPENCV challenges, both times finalists.

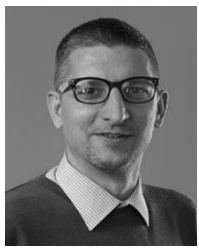

**Vasa M. Radonić** received a Ph.D. degree from the Faculty of Technical Sciences, University of Novi Sad, Novi Sad, Serbia in 2010 in the field of microwave engineering.

He currently works as a Research Professor and Assistant Director for Science at BioSense Institute, Novi Sad, Serbia. He participated in FP6, FP7, H2020, COST, EUREKA, and IPA research projects. He published 3 book chapters and more than 100 papers and serves as a reviewer in several leading scientific journals. His main expertise is microfluidics, bio-sensors development electronic, RF and microwave passive components, fabrication technologies, measurements, and characterization.

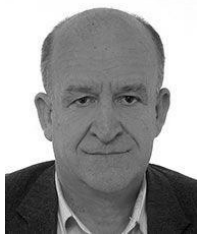

**Ilija D. Brčeski** received a Ph.D. degree from the Faculty of Chemistry, University of Belgrade, Belgrade, Serbia in 2004.

He has been working at the Faculty of Chemistry, Belgrade, Serbia since 1988, currently with the title of Full Professor. He teaches the subjects of Inorganic Chemistry, Natural Resources, as well as Inorganic Syntheses. He is a co-editor of "Journal of Environmental Protection and Ecology", as well

8     IEEE SENSORS JOURNAL, VOL. XX, NO. XX, MONTH X, XXXX

as a member of the editorial board of the publication Archives of Public Health.

Dr. Ilija Brčeski is a member of the European Academy of Sciences and Arts, Salzburg, Austria, and the Serbian Chemical Society.

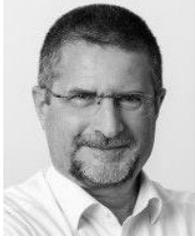

**Rok M. Štravs** received B.S. from the Faculty of Mathematics and Physics, University of Ljubljana, Ljubljana, Slovenia in 1998.

He has more than twenty years of experience running a biotechnological company BIA d.o.o., an established Slovenian supplier of laboratory and process equipment with vast experience in new laboratory technologies, and their application in practice. He is also a leader of the BIA research group Development – Biotechnology and he has been involved in various development projects as a member or team leader. In addition to business and organizational experiences, he has more than forty years of experience in programming control SCADA systems in industry and laboratories and a decade of experience in statistical data processing and Web technologies.

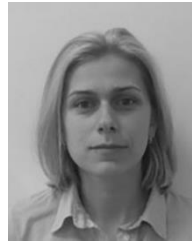

**Nikolina N. Janković** received a Ph.D. degree from the Faculty of Technical Sciences, University of Novi Sad, Novi Sad, Serbia in 2013 in the field of microwave engineering.

She currently works as a Senior Research Associate in the Center for Sensing Technologies at the BioSense Institute, Novi Sad, Serbia. She has authored and co-authored more than 40 journal and conference papers and 5 book chapters. Her main research interests include metamaterials, plasmonics, optics, sensors, and microwave passive devices. She has coordinated H2020 MSCA-RISE project NOCTURNO and EUREKA project WaQUMoS in Serbia. In addition, she has participated in two FP7 projects and four H2020 projects.

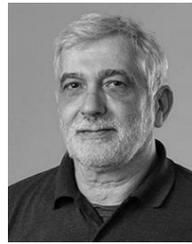

**Jovan B. Matović** received a Ph.D. degree from the Electrotechnical Faculty, University of Belgrade, Belgrade, Serbia in 2002 on the subject of applied physics.

He currently works as a Principal Scientist in the Center for Sensing Technologies at the BioSense Institute, Novi Sad, Serbia. He works in various fields of science and technology, including the development of new semiconductor fabrication processes of high-performance sensors, plasma synthesis of nanoparticles, and the design of satellite thermal control for ESA. Along with instrument conceptualization and design, he also works on novel materials for infrared detectors. He received awards from European Space Agency for novel solution for satellite thermal management in 2004 and a winning Challenge at Nature & Cleveland Clinic, for novel neurosurgery instrument "Avoidance of Blood Vessels During Insertion of Medical Probes" in the amount of 30.000$ in 2012. He has over 1100 and citations and was project manager in multiple Austrian and EU projects i-10 index 25.